\documentclass{article}

\usepackage{PRIMEarxiv}

\usepackage[utf8]{inputenc} 
\usepackage[T1]{fontenc}    
\usepackage{hyperref}       
\usepackage{url}            
\usepackage{booktabs}       
\usepackage{amsfonts}       
\usepackage{nicefrac}       
\usepackage{microtype}      
\usepackage{lipsum}
\usepackage{fancyhdr}       
\usepackage{graphicx}       
\graphicspath{{res/}}     

\usepackage{amsmath,amssymb}
\usepackage{comment}

\title{Vector Symbolic Open Source Information Discovery}

\author{
    Cai Davies \\
    Cardiff University \\ UK
     \And 
    Sam Meek \\
    Helyx Secure Information Systems Ltd \\ UK
     \And 
    Philip Hawkins \\
    Helyx Secure Information Systems Ltd \\ UK
     \And 
    Benomy Tutcher \\
    Frazer-Nash Consultancy \\ UK
     \And 
    Graham Bent \\ 
    Neurosynapse Ltd \\ UK
     \And 
    Alun Preece \\
    Cardiff University \\ UK
}

\begin{document}
\maketitle

\begin{abstract}
Combined, joint, intra-governmental, inter-agency and multinational (CJIIM) operations require rapid data sharing without the bottlenecks of metadata curation and alignment. Curation and alignment is particularly infeasible for external open source information (OSINF), e.g., social media, which has become increasingly valuable in understanding unfolding situations. Large language models (transformers) facilitate semantic data and metadata alignment but are inefficient in CJIIM settings characterised as denied, degraded, intermittent and low bandwidth (DDIL). Vector symbolic architectures (VSA) support semantic information processing using highly compact binary vectors, typically 1-10k bits, suitable in a DDIL setting. We demonstrate a novel integration of transformer models with VSA, combining the power of the former for semantic matching with the compactness and representational structure of the latter. The approach is illustrated via a proof-of-concept OSINF data discovery portal that allows partners in a CJIIM operation to share data sources with minimal metadata curation and low communications bandwidth. This work was carried out as a bridge between previous low technology readiness level (TRL) research and future higher-TRL technology demonstration and deployment.
\end{abstract}


\section{Introduction}

Global data is estimated to reach $163$ zettabytes by 2025, with around $80\%$ being unstructured \cite{rydning2018digitization}, i.e., data lacking metadata either partially or completely; for example, a written report contains the important information in natural language, with the report's metadata such as creation and modification time providing minimal additional context. This data deluge will result in users spending more time finding relevant data than making use of it \cite{seeping_semantics} across all domains including medicine \cite{CALHOUN2014262}, education \cite{data_education}, business \cite{data_business}, and defence.  

In the case of defence, data discovery is a prerequisite for data consumption and exploitation. 
The UK Ministry of Defence (MOD) Data Life Cycle~\cite{MOD:2021} defines four stages: creation, curation, consumption and exploitation,  where discovery is facilitated by curation. Unfortunately, curation is often a significant bottleneck and particularly challenging where a centralised approach is infeasible, notably where data needs to be shared between multiple partners each with independent domains of data control, such as in so-called combined, joint, intra-governmental, inter-agency and multinational (CJIIM) operations. Data and metadata consistency and reuse are much harder to achieve in a CJIIM context due to heterogeneity of data platforms, formats, and standards. Automated approaches to metadata generation and alignment are therefore key. However, a tendency to generate ever-increasing volumes of descriptive metadata only adds to the data deluge.

CJIIM operations can be seen as an instance of any setting where entities share data in a partnership, where standards or policies can differ significantly, and so here is a need for data discovery methods that are independent of a single agreed set of conventions or schemas. From the concept of FAIR Data Principles\footnote{https://force11.org/info/the-fair-data-principles (accessed 23/7/24)}, data must be Findable, Accessible, Interoperable and Re-usable. Data discovery between partners can cause security and privacy concerns, hence a method that allows for access control while maintaining the ability for discovery that does not impact either is crucial.

A further complicating factor in many defence contexts is that data and metadata need to be shared and exploited in a setting characterised as denied, degraded, intermittent and low bandwidth (DDIL). This means that any approaches taken to addressing data curation, discovery, consumption and exploitation need to be highly efficient in terms of communication constraints, especially bandwidth. 

This paper presents a novel approach to addressing the problem of data discovery and alignment in a CJIIM DDIL setting, demonstrated by a proof-of-concept application. 

\section{Technical Approach}

In recent years, embedding models, transformer models, and large language models (LLMs) such as BERT \cite{devlin2019bert}, MPNet \cite{song2020mpnet} and many others have emerged as a promising approach to automating semantic mapping of unstructured and semi-structured data. In a natural language processing (NLP) context, they are used to represent text in a vector form, where computed distances equate to semantic similarity \cite{reimers2019sentencebert}. This method can be used to discover semantically similar data. In contrast to traditional search engines which only find documents based on lexical matches, semantic searches can leverage synonyms and similar contexts. Asymmetric corpora \cite{bajaj2018ms} allow for semantic search on asymmetrically-sized texts; usually a search query is much shorter than the required information to be retrieved. While these approaches are promising for automating data and metadata alignment, the significant size of the vectors (e.g., 768 real numbers for BERT) makes their suitability in a DDIL setting questionable.

Vector symbolic architectures (VSA) \cite{Schlegel_2021} are a form of brain-inspired computing, employing a set of operators in a high-dimensional vector space to represent data and perform symbolic computations. VSA's distributed representations make these computations robust, scalable and efficient for model training and inference. VSA vectors can be real-valued, such as \textit{holographic reduced representations} (HRR) \cite{plate1995holographic} or binary, such as \textit{binary spatter code} (BSC) \cite{kanerva1996binary}. A key advantage of using BSC over HRR is efficiency in both processing --- bitwise XOR instead of discrete-time Fourier transform --- and memory. Projection into binary VSA is a one-way function through a random matrix, offering a form of encryption or obfuscation in that it is computationally infeasible to determine the exact underlying data. Highly-compact binary VSA vectors are promising for use in DDIL settings.

\begin{figure}[ht]
    \centering
    \includegraphics[width=0.6\textwidth]{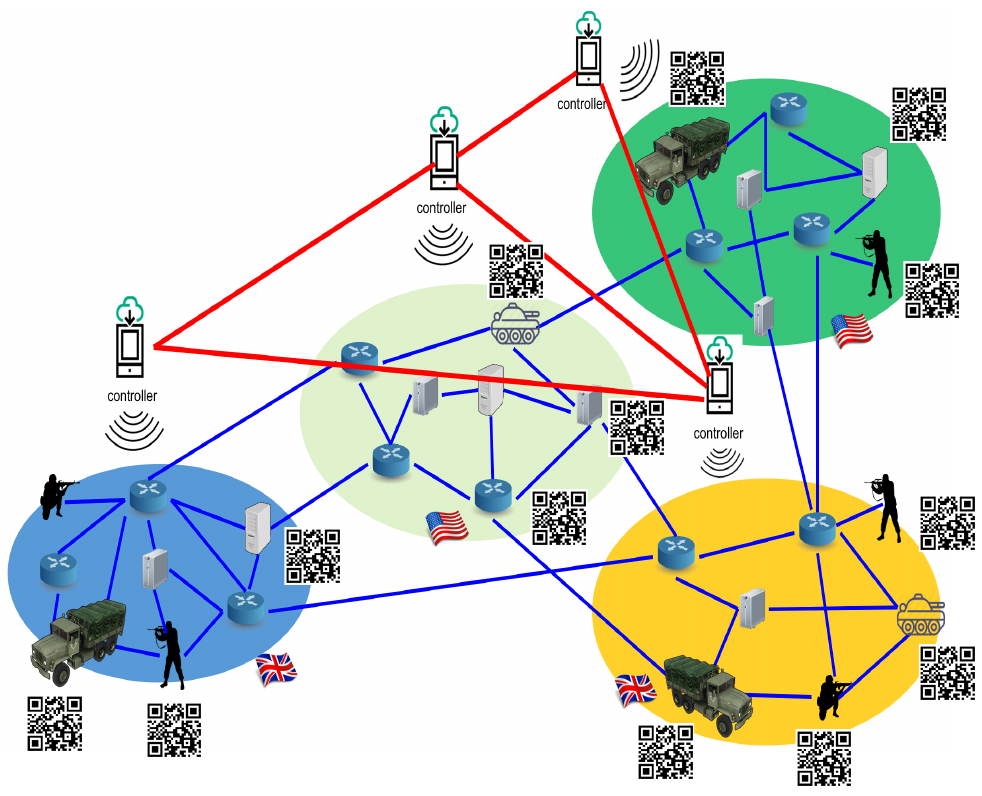}
    \caption{Illustration of DAIS `distributed brain' CJIIM concept (adapted from https://dais-legacy.org/1a11).}
    \label{fig:distributed-brain}
\end{figure}

The application in this paper builds on basic research originating in a joint UK MOD / US Department of Defense programme: Distributed Analytics and Information Science (DAIS)\footnote{https://dais-legacy.org (accessed 23/7/24)}. In the context of future military CJIIM operations, DAIS examined how multiple sensors, devices and services (assets) owned by different coalition partners could be dynamically combined into workflows able to perform new complex tasks on demand. 
DAIS pioneered the use of VSA to create a `distributed brain' for a CJIIM coalition \cite{Simpkin:2019,Simpkin:2020,Verma:2017}: Figure~\ref{fig:distributed-brain}. The `QR code' symbols illustrate the symbolic vectors; these descriptions of coalition sensors and services are generated and stored locally. The required configuration of assets is constructed as a vector of vectors and transmitted over the DDIL communications network in a highly compact 1-10k bit format. Through a process of peer-to-peer vector exchange the required assets are discovered and linked together to perform the required task~\cite{Simpkin:2020}. The process of generating the compact binary vectors is inherently lossy, therefore, the vectors need to be matched with `clean' copies from a (local) dictionary to resolve matches accurately.

The application in this paper extends the DAIS use of hyperdimensional vector representations to data discovery and matching, in both semantic and structured data representations. A key concept is the ability of VSA to represent multiple attributes in a single vector space by means of projection, binding and bundling (see Methods). 
As an efficient heuristic matching method, we make use of FAISS \cite{johnson2017billionscale} for high-dimensional vector indexing and matching, which improves over XOR operations on conventional hardware. We identify open-source information (OSINF) as a setting for our application, and Twitter as the data source, with the goal to allow an analyst to discover data for a given task through a combination of semantic similarity and structured metadata matching, in a way that is compact and efficient for use in DDIL environments and low-resource devices, allows for independence of conventions or schema, and allows for data discovery without compromising privacy and access control through the inherent encryption/obfuscation offered by VSA.

The aim of this work is therefore to demonstrate that defence can exploit semantic vectors and VSA to enhance interoperability of metadata to support data discovery between CJIIM partners. Specifically, the work aims to advance the DAIS VSA approach from low technology readiness level (TRL) to proof-of-concept application.

\section{Related Work}

There is very little prior work on the integration of transformer models and VSA or hyperdimensional computing, confirmed by a recent survey~\cite{Kleyko:2023} in which only a single paper \cite{ma_and_ma} --- discussed below --- combines a transformer method with a real number vector hyperdimensional computing approach.

One approach\cite{seeping_semantics}, like our work, uses NLP embedding models to discover semantically related objects. However, they use word, rather than sentence embeddings, and use traditional knowledge graphs rather than VSA approaches to structured knowledge encoding.

The approach mentioned above \cite{ma_and_ma} combines multi-modal transformer-based models with hyperdimensional computing methods, namely HRR, to capture high-order dependency with a significant reduction in computation by making use of the compressed representation. These transformer-based models differ from large language models both in terms of the architecture, complexity, and captured natural language capabilities.


Therefore, while ours is not the first approach to make use of semantic similarity for data discovery, or the use of transformer-based architectures in high-dimensional computing, we believe our work is the first integration of transformer models with VSA, combining the power of transformers for semantic matching with the compactness and representational structure (binding and bundling) of VSA.

\section{Problem Example: Open Source Information}

It is now widely accepted that information provided from open sources --- open-source information (OSINF) --- including social media, can give broad insight into aspects of an unfolding situation~\cite{Olcott:2012}. 
OSINF is particularly challenging from the perspective of data discovery for a variety of reasons. First, the sources are external, not under CJIIM control to curate. Secondly, the space of sources is dynamic --- platforms come and go with regularity --- and populations of interest tend to migrate from one platform to another. Thirdly, content is semi-structured, with much of the exploitation value in the unstructured content (e.g., text, still images, video), so approaches to discovery need to encompass that unstructured material.

One of the most prominent social media platforms of the past decade, Twitter, also illustrates the notion of metadata explosion. Twitter's application programming interface (API) documentation  indicates that over 300 metadata fields may be attached to a single tweet.\footnote{https://developer.twitter.com/en/docs/twitter-api (accessed 23/7/24)} 
An initial exploration of the dataset described in Methods below showed that there is approximately a 10:1 ratio of metadata to data in these tweets by volume (where \textit{data} is defined as those fields forming the displayed content of the tweet, mainly the tweet text).

The proof-of-concept application implements  \textit{Speeding-up the OSINF request-for-information (RFI) process} as a generic use case that maps widely across many domains. The manual RFI process typically proceeds as follows:
\begin{enumerate}
\item Requirement is formulated for OSINF data;
\item Request is sent to data manager;
\item Data is manually gathered by a data manager --- this often involves re-collecting data already available somewhere but not properly curated;
\item Data is sent back to requester.
\end{enumerate}

It is not uncommon for this process --- involving significant human effort --- to take considerable time (weeks). In a CJIIM setting, step~3 is significantly more difficult as the multiplicity of sources and high variety of data formats and inconsistent levels of curation is significantly greater.

\section{Proof-of-Concept Approach}

\begin{figure}[ht]
    \centering
    \includegraphics[width=0.8\textwidth]{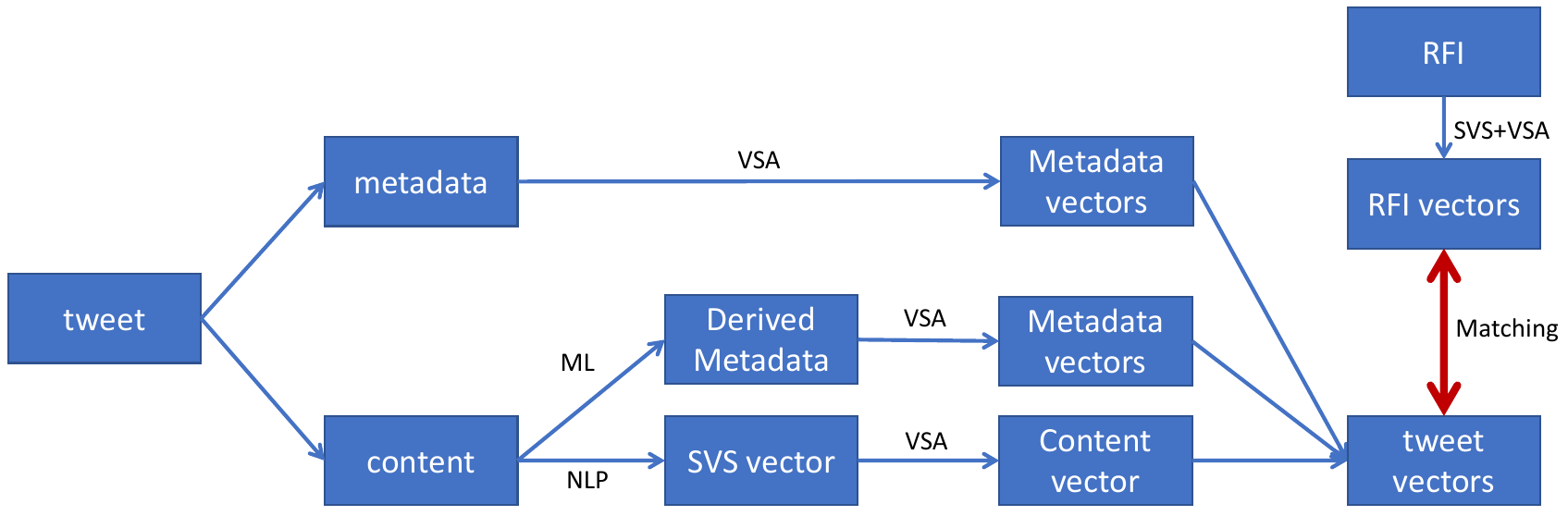}
    \caption{Overview of proof-of-concept (Twitter) approach.}
    \label{fig:approach-overview}
\end{figure}

An overview of the approach for the OSINF data discovery proof-of-concept is shown in Figure~\ref{fig:approach-overview}. Tweet metadata attributes are mapped via VSA operators into metadata binary vectors. Some of these attributes such as timestamp are extracted directly from the original tweet metadata while others, such as sentiment or location, are derived from the tweet content by machine learning (ML) services. Tweet semantic content is first mapped to real-numbered vectors via NLP (LLM) and then to VSA binary vectors. The final vector representation of a tweet is created from these component metadata and content VSA vectors. For matching, a user's RFI is also encoded via a combination of the LLM and VSA methods. Matching between RFI and tweets takes place in the VSA vector space.

\begin{figure}[ht]
    \centering
    \includegraphics[width=0.6\textwidth]{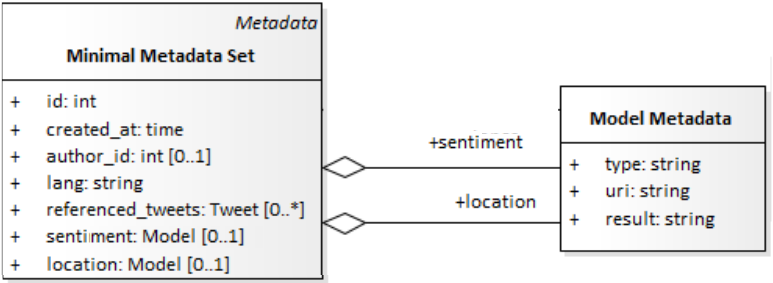}
    \caption{Minimal metadata model (UML schema).}
    \label{fig:metadata_model}
\end{figure}

Metadata models for VSA must be structured and somewhat semantically interoperable to ensure the searches and vector comparisons are appropriate. Our approach is to create a \textit{minimal metadata model}, Figure~\ref{fig:metadata_model}, intended to be generic across multiple social media platforms. The model includes ML-derived attributes such as sentiment and location, important in the OSINF context. 

In the proof-of-concept application, Twitter was used as the sole source for social media data with a view to identifying similar tweets from hashtags and target examples. It is however entirely feasible to utilise multiple data types and social media feeds in the Figure~\ref{fig:approach-overview} process. For example, it should be possible to use a tweet as an input to the RFI application and retrieve different data sources based upon similarity, e.g., Reddit posts, Telegram messages, or TikTok comments.

\section{Methods}

\subsection{Data}
The Twitter API 
was used to collect approximately 50,000 anonymized tweets (data and metadata) with given hashtags spanning multiple topic areas, including Russia/Ukraine, the stock market, a combat video game (Warthunder) and others. These were selected for both semantic diversity and potential overlap, e.g., we would expect some overlap of tweets on war-based video games with real-world war content, but little to no overlap between war and sports.

\#acecombat, \#art, \#belarus2020, \#belarusprotests, \#covid19, \#dcsworld, \#donetsk, \#freebelarus, \#gaming, \#jobs, \#kazakhstanprotests, \#kherson, \#luhansk, \#nhl, \#rugby, \#standwithbelarus, \#standwithputin, \#standwithrussia, \#standwithukraine, \#stocks, \#stoprussia, \#supportbelarus, \#supportrussia, \#supportukraine, \#vegan, \#warthunder, \#zaporizhzhia.


\subsection{Semantic Embedding}
The SentenceTransformers library \cite{reimers2019sentencebert} provides a framework for computing sentence/text embeddings with a given model; for this work we chose `all-mpnet-base-v2', but many alternatives exist. Our goal for the VSA approach is matching in a convention-agnostic vector space, where the choice of embedding models is not limited; i.e., CJIIM partner A can use a different embedding model to partner B and still match in a shared vector space. We use the model to generate semantic vector representations for the text and hashtags of our collected set of tweets. 

\subsection{Vector Symbolic Architecture}
Here we summarise VSA operations used with binary vectors (BSC) \cite{kanerva1996binary}. The terms high-dimensional vector and hypervector are used interchangeably. Common distance measurements in VSA are cosine-distance for real-valued vectors and Hamming distance for binary vectors.

\textbf{Projection} transforms real-valued vectors into a high-dimensional VSA space. For BSC, we create a random binary matrix $\textbf{B}^{b\times r}$ which is our projection matrix, and we create the binary vector $\textbf{z}$ from real-valued vector $\textbf{r}$:
\begin{equation*}
\begin{split}
    \textbf{A} & = \textbf{B} \times \textbf{r} \\
    \textbf{a}_{b} & = \sum^{r}_{i=1} \textbf{A}_{bi} \\
    \textbf{z} & = \frac{\textbf{a} - \mathbb{E}(\textbf{r})}{\textrm{Var(\textbf{r})}} \\
    \textbf{z}_{i} & = 
    \begin{cases}
        1, & \textbf{z}_{i} \geq 0 \\
        0, & \text{otherwise}
    \end{cases}
\end{split}
\end{equation*}

\textbf{Bundling} is a method of creating a `compound' vector (vector of vectors) by a majority-sum operation. With a list of vectors $[\textbf{v}_{1}..\textbf{v}_{n}]$ as matrix $\textbf{V}^{i\times n}$, we bundle to create a compound vector $\textbf{z}$:
\begin{equation*}
\begin{split}
\textbf{z}'_{i} & = \frac{\sum_{j=1}^{n} \textbf{V}_{i,j}}{n} \\
\textbf{z}_{i} & = 
    \begin{cases}
        1, & if \: \textbf{z}'_{i} > 0.5 \\
        0, & if \: \textbf{z}'_{i} < 0.5 \\
        \textit{random} & \text{otherwise}
    \end{cases}
\end{split}
\end{equation*}
Each tweet can thus be represented as a single vector (a compound of its data and metadata attributes) which allows comparison of single vectors rather than per-attribute matching:
\begin{align*}
    Tweet_{v} = [text_{v} + hashtags_{v} + ... + created_{v}]
\end{align*}
Given the distributed nature of VSA vectors, this allows the storing of information of multiple vectors into a single vector, but at the cost of introducing noise; there is a trade-off between the size of the vector, the number of vectors to bundle, and the threshold of vector degradation.

\textbf{Basis vectors} are random hypervectors used to represent the smallest granular units of information, e.g., for text, letters of the alphabet and common symbols. To create words, we bundle a set of hypervectors together, e.g.:
\begin{align*}
    hello = [h_{v} + e_{v} + l_{v} + l_{v} + o_{v}]
\end{align*}
We can hierarchically bundle words to create sentences, sentences to paragraphs and so on, with the ability then to calculate distances between entire documents by the top-level vector. Random hypervectors are close to orthogonal, forming a tight distribution around a Hamming distance of $0.5$; for 10k bit binary vectors, fewer than $1$ in $10^9$ of any two vectors will have a Hamming distance less than $0.47$.

\textbf{Level hypervectors} are a sequence of basis vectors $[L_{1}...L_{m}]$ where each element in the sequence has a linearly increasing Hamming distance from $L_{1}$ to $L_{m}$, such that $L_{1}$ and $L_{m}$ are orthogonal with a Hamming distance of $0.5$; this allows representation of linear relationships such as time.

\textbf{Binding} is an XOR operation between two vectors, and acts as key-value assignment. The `role' vector, a random hypervector, acts as the key, and the `filler' (data) vector acts as the value. For example, if $Z=X\cdot A$, then $X\cdot Z = X\cdot (X\cdot A) = A$ where $\cdot$ is the XOR operation. Due to the distributive property of VSA, we can extract noisy versions of filler vectors from compound vectors by means of XOR. For example, we consider as filler vectors $A$ and $B$, with $X$ and $Y$ representing the role vectors:
\begin{equation*}
\begin{split}
    Z & = [X\cdot A + Y\cdot B] \\
    A' & = Z\cdot X
\end{split}
\end{equation*}
For OSINF we create role hypervectors for each tweet attribute. 
By creating the role vectors from level hypervectors and using the element in the middle of the sequence, which has a Hamming distance of $0.5$ to $L_{1}$, we can create a weighting of attributes in the query vector (see Matching).

\subsection{OSINF Metadata Representation}

\textbf{Tweet text and hashtags} are both represented semantically, making use of LLMs to create semantic vectors representing the text, which are then transformed into the VSA space by means of projection. For tweet text we create a single semantic vector which is then projected into VSA; for hashtags we create a semantic vector for each hashtag in the tweet, project each into the VSA space, then bundle the vectors together to create a compound hashtag vector. Projecting the LLM-derived semantic vectors (768 real numbers) to 1k bit VSA vectors results in a $25 \times$ compression.

\textbf{Location and Language} are represented lexically, so that the resulting vectors will match most closely with text spelt similarly. To achieve this, \textit{basis vectors} are created for each character or symbol, and bundled together to create a compound vector representing the text. Thus, `en-uk' and `en-us' are similar but distinct vectors. To extract location, we use NER\footnote{https://huggingface.co/Davlan/bert-base-multilingual-cased-ner-hrl (accessed 23/7/24)} to find instances of named locations in the tweet text. For simplicity, we assign the first instance to be the location.

\textbf{Sentiment} is first derived from tweet text by means of a sentiment classification model, e.g., `cardiffnlp/twitter-roberta-base-sentiment-latest'\footnote{https://huggingface.co/cardiffnlp/twitter-roberta-base-sentiment-latest (accessed 23/7/24)}; with this model, the result is a 3-dimensional vector representing probabilities of negative, neutral and positive sentiment. This vector can then be transformed into VSA space using projection as above. This representation can be generalised  any ML-derived results of any vector size.

\textbf{Timestamps} (created\_at) are represented using level hypervectors, where each element in the sequence is a representation of a time window, with $L_{1}$ and $L_{m}$ representing the start and end time; similar to UNIX time representing the number of seconds since 1/1/1970. The trade-off for the resolution of the time windows is determined by size of the time range to represent and the size of the hypervectors. An increase in vector size allows finer resolution time windows, whereas an increase in the represented time range  decreases the resolution. An alternative is to split the timestamp into constituent parts (year, month, day, etc), and treat each as a separate attribute. Then the basis vectors method can be used to map each possible value in year $\{1970..2100\}$, month $\{1..12\}$, day $\{1..31\}$, etc, to a random hypervector, such that every value is orthogonal to every other value.

\subsection{Matching}

\begin{figure}[ht]
    \centering
    \includegraphics[width=0.8\textwidth]{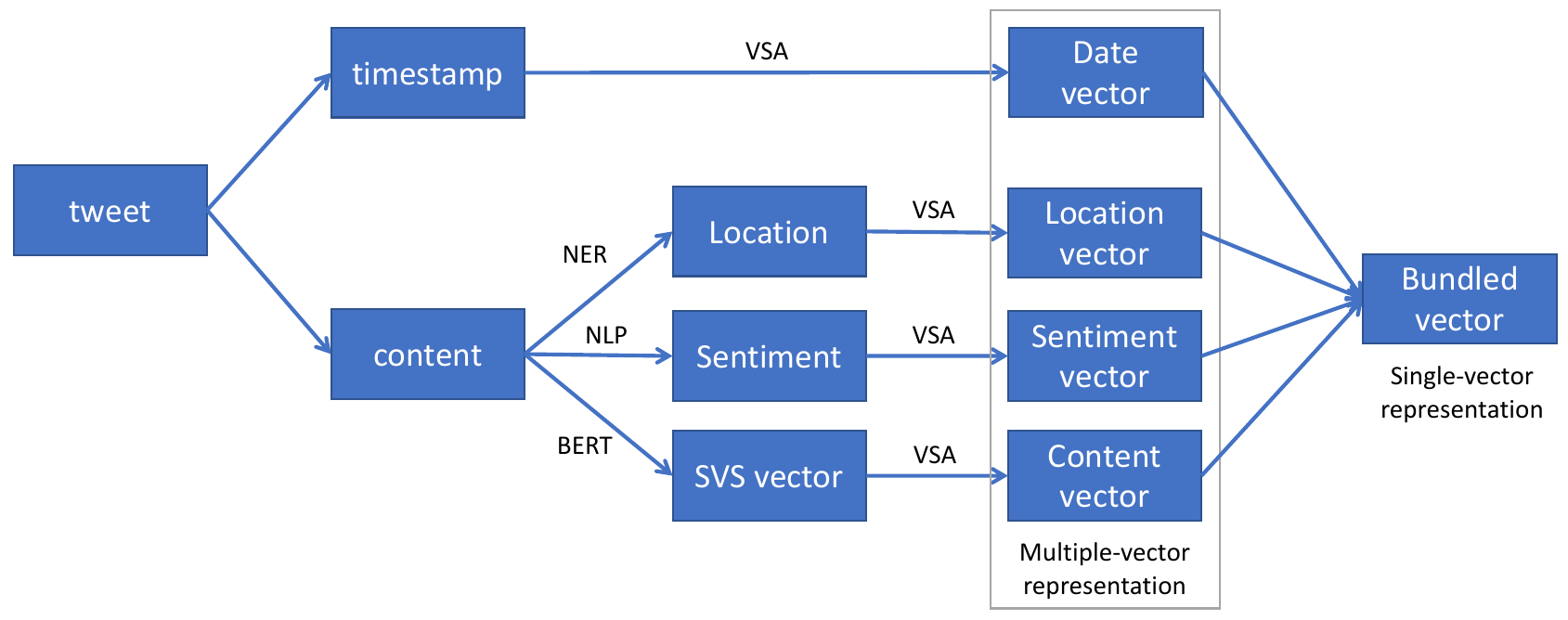}
    \caption{Method for representing tweets as vectors.}
    \label{fig:approach-detail}
\end{figure}

As described above, we can represent a tweet by two methods: (1) \textit{multiple-vector}, one vector for each attribute; (2) \textit{single-vector}, by first binding each attribute `filler' vector to a created `role' vector (selected from the middle of separate level hypervector sequences) and then bundling into a single compound vector. This workflow is illustrated in Figure \ref{fig:approach-detail}.

The query is represented in VSA using the same methods as described above, which can be multiple-vector or single-vector; the difference for a single-vector query being the creation of a vector for each attribute in the query with all `role' vectors weighted to $0$ except the `role' for the given attribute; such that for $N$ attributes to match, we have $N$ query vectors. From this, we can then calculate the Hamming distance for each attribute across all tweets, which gives matrix $D^{a\times t}$ where $a$ is the number of attributes in the query, and $t$ the number of tweets in the set. The benefit of single-vector is seen in storage and memory; rather than $a\times t\times b$ vectors where $b$ is the VSA dimension, we have $t\times b$ vectors to represent the set of tweets in VSA.

Alongside the query, we can pass parameters for thresholding the Hamming distance, which we call `fuzziness'; we pass a threshold for each attribute in the query. Combined with the distance matrix $D$ we create a mask, where for each attribute $a$ the element $D_{a,t}$ is true only if less than the associated fuzziness value. We then take the intersection across the attributes to give the resulting set of matched tweets.

This brute-force approach is inefficient, and so we make use of FAISS \cite{johnson2017billionscale} for vector matching in the binary space through indexing. For multiple-vector, we have $a$ indices, and for single-vector a single index. Given a query vector, FAISS returns the top $K$ distances and indices, which can be used as above with the fuzziness parameters and intersections to return a set of matched tweets. Issues here arise with the value of $K$; there is a linear increase in search time with FAISS, and an increase in time for computing threshold maskings and intersections, but the correct value of $K$ for the given fuzziness parameters is unknown. Our approach sets $K$ to the number of tweets in the whole set, but for scalability this approach should be re-examined.

\section{Results and Discussion}
Here we summarise experiments to assess matching accuracy 
in multiple-vector and single-vector representations, using 1k and 10k bit VSA vector size. 

\textbf{Tweet text and hashtags:}
To assess matching these semantic representations it was necessary to manually label matches for queries as on-topic or off-topic. We created three topics by randomly selecting a seed tweet in each case that was deemed an exemplar of a topic: `Russia-Ukraine', `Stocks' and `Warthunder' (combat video game) respectively. These three topics were chosen for their broad semantic separation, as well as some anticipated semantic overlap between Russia-Ukraine and Warthunder. For each of these randomly-selected seeds, we took the top $300$ matches and labelled them as on- or off-topic; accuracy is then the proportion of on-topic matches. Table \ref{tab:sem_results} shows the results for multiple-vector (MV) 1k bit and single-vector (SV) for 1k and 10k bits.

\begin{table}[ht]
\centering
\begin{tabular}{|l|l|l|l|}
\hline
\textbf{Topic} & \textbf{MV 1k} & \textbf{SV 1k} & \textbf{SV 10k} \\ \hline
Russia-Ukraine & 1.0                      & 0.883                     & 1.0                        \\ \hline
Stocks         & 0.993                    & 0.8                       & 0.973                      \\ \hline
Warthunder     & 0.997                    & 0.52                      & 0.983                      \\ \hline
\end{tabular}
\caption{Semantic matching results}
\label{tab:sem_results}
\end{table}

\textbf{Location and Language:} For these lexical representations accuracy is a measure of recall when dividing the matches (in order) into two sets at the point of the known ground-truth; e.g., if the query is `en-uk', and it is known that $x$ tweets are `en-uk', then we classify matches (in order of Hamming distance) as positive for the first $x$ and negative for the rest. From this, we can measure the recall, given we have ground-truth and `classified' labels. For multiple- and single-vector, this resulted in recall of $100\%$ for both 1k and 10k bit representations across all possible values for location and language in our set of 50K tweets.

\textbf{Sentiment:} We compute accuracy as above, where we first take the argmax of the 3-dimensional vectors as the `class', as well as for the query (which we define as the one-hot vector of the chosen class); then, positive classes are those that match the query, negative for classes other than the query class. As above, we use the known $x$ tweets with ground-truth equal to the query class as predicted positive, and compute the recall; this is done for each possible class of negative, neutral and positive. For multiple-vector in 1k bit representation, we see $97.84\%$ accuracy, but this drops to $84.85\%$ in single-vector 1k bit representation. 10k bit shows an accuracy of $97.84\%$ for both vector sizes.

\textbf{Timestamp:} Represented using level hypervectors, we consider the order of retrieved matches with a query vector representing the earliest possible time, such that the order should be temporal. Given that a usual accuracy of this ordering is determined by the resolution of the time window, which can be determined by any number of combinations of time range and vector size, we simply consider a time to be `correct' if the order of the matching ensures it is in the correct time window. For multiple-vector, this method gives $100\%$ accuracy; however for single-vector we see a significant degradation in accuracy to close to random sampling, which is due to noise introduced by bundling. Using the basis vectors method, we achieve $100\%$ in single-vector, at the cost of additional matching operations if given a time range (i.e., for year, month, day).

\subsection{Discussion}
These results indicate that the combination of LLM and VSA provides effective approaches to data discovery in the OSINF setting. The LLM to VSA mapping (up to $25 \times$ compression) preserves semantic matching well, and the VSA binary vector encoding provides a highly compact (1k bit) vector representation suitable in a DDIL environment. These findings are important because OSINF represents a particularly challenging metadata problem space, as detailed in the Introduction. Therefore, we would expect the results to generalise to other, more controlled and curated data sources across a range of application domains.

The multiple-vector representation offers the better trade-off between accuracy and compactness, compared to the single-vector representation. Three `dimensions' --- temporal, spatial, and semantic --- plus sentiment can be captured in 4k bits (small enough to fit in a minimum size IPv4 datagram) with good matching accuracy across all attributes. If single-vector is preferred, for example to allow for a larger set of attributes, then the 10k bit setting is preferable to the 1k bit encoding. Other options could include combinations of multiple- and single-vector, e.g., to separate `high resolution' representations such as timestamps into a separate vector, and bundle other attributes into a compound vector.

\section{Proof-of-Concept Application}

\begin{figure}[t]
    \centering
    \includegraphics[width=0.67\textwidth]{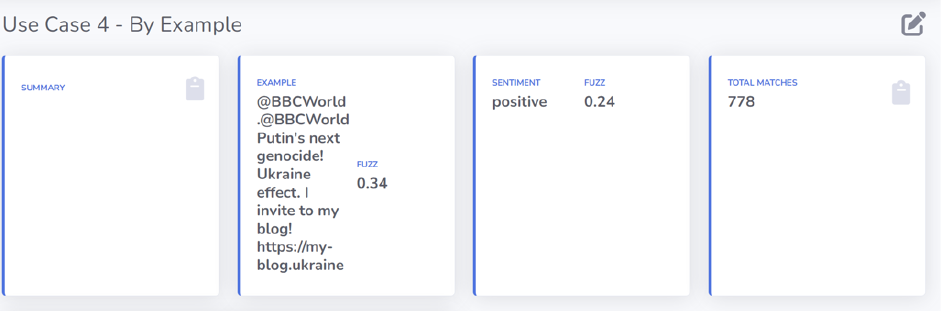}
    \caption{Proof-of-concept app: query-by-example.}
    \label{fig:poc-dashboard}
\end{figure}

\begin{figure}[t]
    \centering
    \includegraphics[width=0.33\textwidth]{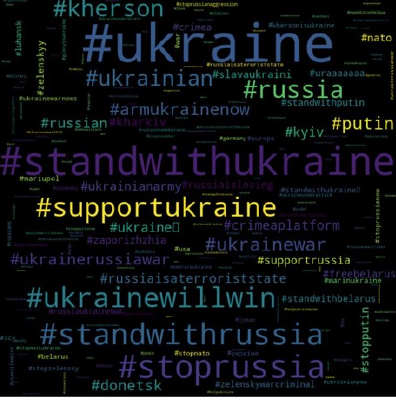}\includegraphics[width=0.5\textwidth]{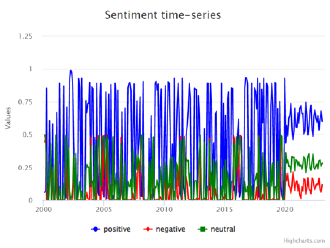}
    \caption{Illustrative word cloud (left) and sentiment (right) of matched tweets.}
    \label{fig:poc-word-cloud-and-sentiment}
\end{figure}

The OSINF Discovery Portal proof-of-concept embodies the above methods and allows RFIs to be built from a flexible palette of elements including time range, geospatial area, search terms, or query-by-example (tweet). Figure~\ref{fig:poc-dashboard} shows the latter, where the user is seeking tweets semantically similar to the specified (synthetic) tweet.
A visualisation of the text of matched tweets is given in the form of a word cloud --- Figure~\ref{fig:poc-word-cloud-and-sentiment} (left) provides illustrative examples --- where it can quickly be seen verified whether the matched content is broadly `on topic'. Other visualisations based on the VSA-encoded metadata include tweet volume and sentiment over time;  Figure~\ref{fig:poc-word-cloud-and-sentiment} illustrates these features. 

\section{Conclusion and Deployment}

This work has demonstrated a novel approach to metadata management and semantic alignment, lowering human, computation and communication burdens, enabling integration and interoperability of data in CJIIM and DDIL settings. The approach has broader applicability in data discovery and sharing across public and private sector domains such as healthcare, education and business. The work has advanced the TRL of prior VSA research. It was undertaken in response to enduring defence capability challenges~\cite{MOD:2020}: (a) pervasive, full spectrum, multi domain intelligence, surveillance and reconnaissance (ISR), (b) multi-domain command and control, communications and computers (C4), and (c) secure and sustain advantage in the sub-threshold. The applied outcomes of this work are now being down-streamed into multiple follow-on taskings. 

\section{Acknowledgements}

This work was funded by the UK Defence Science and Technology Laboratory (Dstl).

\bibliographystyle{unsrt}  
\bibliography{bib}

\end{document}